\documentclass{webofc}

\usepackage[varg]{txfonts}   
\usepackage{graphicx}
\usepackage{subcaption}
\usepackage{hyperref}
\usepackage{url}

\hypersetup{colorlinks=true,citecolor=blue,urlcolor=blue,linkcolor=blue}

\begin{document}
\title{Exploring the Origin of Anisotropy in Small Systems: From Symmetric (O+O) to Asymmetric (d+Au) Collisions}

\author{\firstname{Zhengxi} \lastname{Yan}\inst{1}\fnsep\thanks{\email{zhengxi.yan@stonybrook.edu}} (for the STAR Collaboration)}

\institute{Stony Brook University, Stony Brook, NY 11794, USA}

\abstract{
This contribution reports STAR measurements of azimuthal anisotropies in produced particle distributions of the d+Au and $^{16}$O+$^{16}$O collisions at $\sqrt{s_{NN}} = 200$ GeV, probing the origin of collectivity in small systems. We test the hydrodynamic response of the produced medium by comparing these systems with vastly different initial geometries. The measured elliptic ($v_2$) and triangular ($v_3$) anisotropies, and their event-by-event fluctuations, scale robustly with initial-state eccentricities, consistent with the hydrodynamic behavior expected from a Quark Gluon Plasma (QGP) droplet formed in these collisions. The result also helps constrain the role of sub-nucleon fluctuations in determining the initial conditions. Second, the wide pseudorapidity coverage of the STAR detector is used to investigate longitudinal dynamics. Correlations across different rapidity gaps show no significant impact from flow decorrelation, but non-flow contributions are substantial and require careful subtraction.
}
\maketitle
\section{Introduction}
\label{intro}
A major focus in the study of relativistic nucleus-nucleus collisions is to understand the origin of collective flow in small systems. A common approach aims to distinguish between initial-state momentum correlations and final-state response to the collision geometry by systematically varying that geometry. Previous STAR measurements in p+Au, d+Au, and $^{3}$He+Au collisions revealed a similar magnitude of triangular flow ($v_3$) in the systems~\cite{starcollaborationMeasurementsEllipticTriangular2023}, but the model prediction for the ordering also depends on sub-nucleon scale fluctuations or effects from longitudinal flow decorrelation~\cite{zhao3DStructureAnisotropic2023}. To provide a more controlled test, this work compares new measurements from asymmetric d+Au and symmetric $^{16}$O+$^{16}$O collisions at $\sqrt{s_{NN}} = 200$ GeV. These systems offer a much larger lever arm to test the final-state response, as they produce comparable particle multiplicities from vastly different initial geometries. The wide acceptance of the STAR detectors is also utilized to investigate the role of longitudinal flow decorrelation raised by previous experimental results~\cite{starcollaborationMeasurementFlowCoefficients2024}.

\section{Experimental Details and Analysis Methods}
\label{sec-methods}
The analysis uses data from d+Au and $^{16}$O+$^{16}$O collisions at $\sqrt{s_{NN}} = 200$ GeV collected by the STAR experiment in 2021. Time Projection Chamber (TPC) covers a pseudorapidity range of $|\eta| < 1.5$ and the Event Plane Detectors (EPD) cover $2.1 < |\eta| < 5.1$. 

\subsection{Two-Particle Correlations} \label{subsec-2pc} The collective response to the initial geometry is investigated using two-particle correlations for charged particles with $0.2 < p_T < 2.0$ GeV/c. Azimuthal correlation functions are constructed within the TPC acceptance ($|\eta|<1.5$), with a pseudorapidity gap of $|\Delta\eta|>1.0$ imposed to suppress short-range non-flow. The correlation functions are decomposed into a Fourier series to extract the harmonic coefficients $c_n(p_{\mathrm{T}}^{\mathrm{t}}, p_{\mathrm{T}}^{\mathrm{a}})$. The raw $c_n$ coefficients contain residual long-range non-flow. This contribution is estimated using peripheral collisions and subtracted to obtain the corrected coefficients, $c_n^{\mathrm{sub}}$. The genuine pT-differential flow harmonics are then calculated as:
\begin{equation}
v_n(p_{\mathrm{T}}^{\mathrm{t}})=\frac{c_n^{\mathrm{sub}}(p_{\mathrm{T}}^{\mathrm{t}}, p_{\mathrm{T}}^{\mathrm{a}})}{\sqrt{c_n^{\mathrm{sub}}(p_{\mathrm{T}}^{\mathrm{a}}, p_{\mathrm{T}}^{\mathrm{a}})}}
\end{equation}
The integrated flow harmonics, $v_n\{2\}$, are computed from the subtracted coefficients, reflecting the genuine collective behavior of the system.

\subsection{Subevent Method}
\label{subsec-3sub}
To investigate longitudinal flow decorrelations, we employ the subevent method. We define the pseudorapidity regions (subevents): $\eta_a < |1.0|$, $-5.1<\eta_b< -2.1$, $-1.5<\eta_c<-0.5$, and $0.5<\eta_d<1.5$. The $p_T$-differential harmonics are then calculated from flow vectors as:
\begin{equation}
v_n( p_T) = \frac{\langle q_n^{a}(p_T) Q_n^{*b} \rangle}{\sqrt{\frac{\langle Q_n^{b} Q_n^{*c} \rangle \langle Q_n^{b} Q_n^{*d} \rangle}{\langle Q_n^{c} Q_n^{*d} \rangle}}}
\end{equation}
 The numerator represents the correlation between particles at mid-rapidity and the forward event plane. The denominator, representing the correlation between the two mid-rapidity subevents and the forward event plane, is equivalent to the event plane resolution correction. Non-flow subtraction is done for all four pairs of correlations. Flow decorrelation effects are tested by comparing the $v_n$ from this method (mid-forward correlation) to the standard two-particle method (mid-mid correlation).

\subsection{Four-Particle Cumulants} \label{subsec-4pc} Flow fluctuations are quantified using the four-particle cumulant $v_2\{4\}$. For d+Au collisions, the standard full-event method with particles from $|\eta|<1.5$ is used \cite{atlascollaborationMeasurementLongrangeMultiparticle2018}. For O+O collisions, due to smaller flow signal, the two-subevent method with subevent ranges of $-1.5 < \eta < -0.1$ and $0.1 < \eta < 1.5$ is employed to better suppress non-flow \cite{atlascollaborationMeasurementLongrangeMultiparticle2018}. 

\section{Results and Discussion}
\label{sec-results}

\begin{figure*}
\centering
\includegraphics[width=0.9\textwidth]{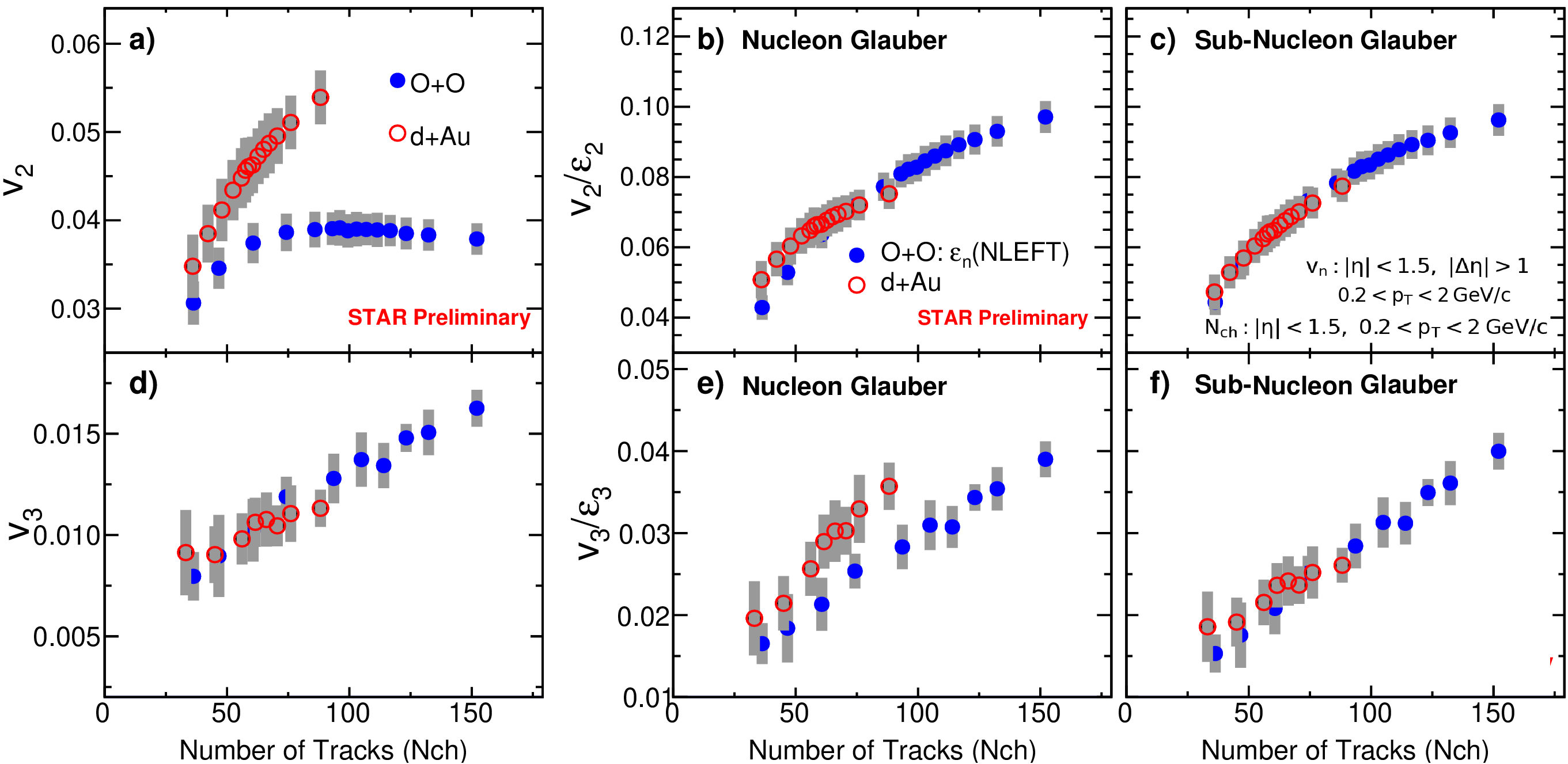}
\caption{Multiplicity dependence of elliptic ($v_2$) and triangular ($v_3$) flow harmonics and their scaling with initial-state eccentricity ($\varepsilon_n$). Panels (a) and (d) show the measured $v_2\{2\}$ and $v_3\{2\}$. Panels (b) and (e) show the scaling ratio $v_n\{2\}/\varepsilon_n\{2\}$ using eccentricities from a nucleon-Glauber model, while panels (c) and (f) use a sub-nucleon (quark) Glauber model~\cite{loizidesGlauberModelingHighenergy2016}.}
\label{fig:v2v3_scaling}
\end{figure*}

\begin{figure*}
\centering
\includegraphics[width=0.8\textwidth]{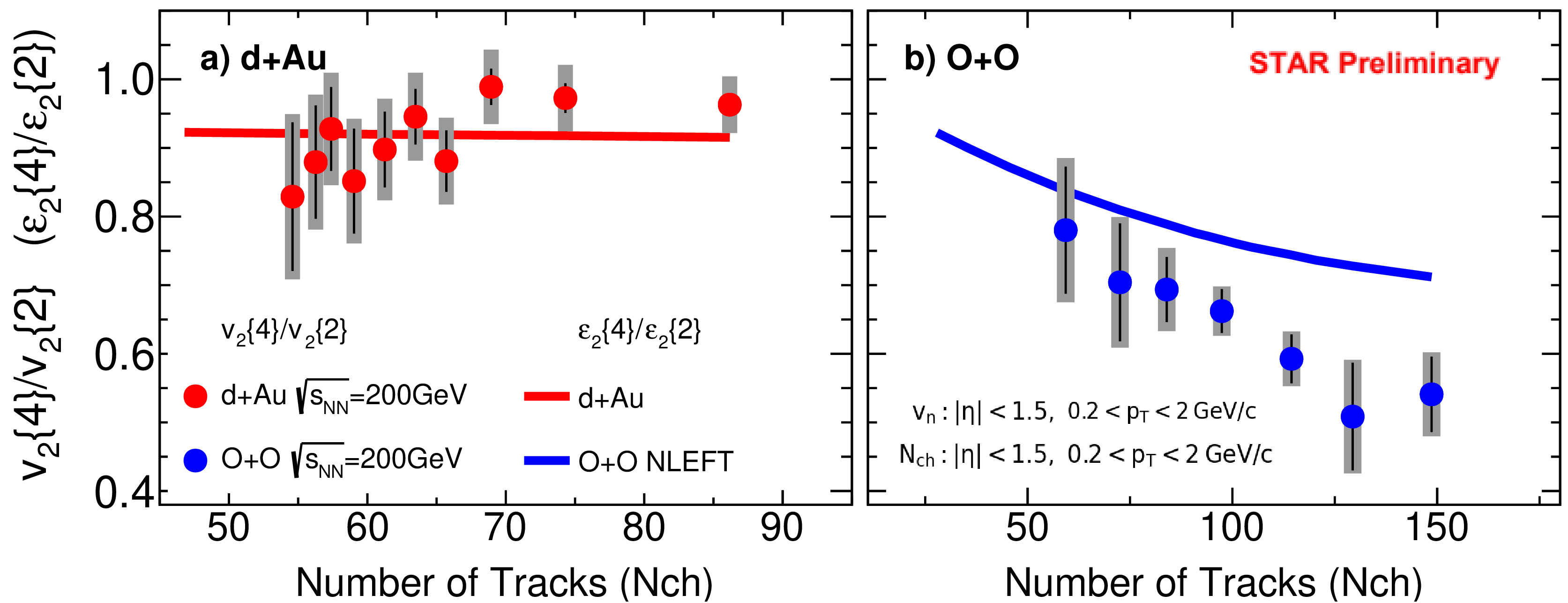}
\caption{Comparison of elliptic flow fluctuations ($v_2\{4\}/v_2\{2\}$, points) with initial eccentricity fluctuations ($\varepsilon_2\{4\}/\varepsilon_2\{2\}$, bands~\cite{luEssentialElementsNuclear2019a}) in d+Au (left) and O+O (right) collisions.}
\label{fig:v24_v22}
\end{figure*}

After non-flow subtraction, the measured flow harmonics in Figure~\ref{fig:v2v3_scaling}(a, d) show a clear hierarchy: $v_2(\text{d+Au}) > v_2(\text{O+O})$, whereas $v_3(\text{d+Au}) \approx v_3(\text{O+O})$. This result directly follows expectations from the initial collision geometry. The larger $v_2$ in d+Au is driven by the significant intrinsic ellipticity ($\varepsilon_2$) from the deuteron's shape. In contrast, the similar $v_3$ values suggest that the fluctuation-driven triangularity ($\varepsilon_3$) is comparable in both systems, which is consistent with models that include sub-nucleon fluctuations~\cite{loizidesGlauberModelingHighenergy2016}. The system's collective response is further quantified by the scaling ratio $v_n/\varepsilon_n$. As shown in Figure~\ref{fig:v2v3_scaling}(b,c,e,f), this ratio is consistent between the two systems for both $v_2$ and $v_3$ across the full multiplicity range. This demonstrates a common response mechanism that converts initial geometry into final-state flow, a feature of QGP droplet formation. The scaling of $v_3$ is better described when $\varepsilon_3$ is calculated with sub-nucleon fluctuations. The response to geometric fluctuations was tested using higher-order cumulants~\cite{bhaleraoEccentricityFluctuationsElliptic2006}. The measured flow fluctuation ratio $v_2\{4\}/v_2\{2\}$ scales with the initial eccentricity fluctuation ratio $\varepsilon_2\{4\}/\varepsilon_2\{2\}$, as shown in Figure~\ref{fig:v24_v22}, further supporting the geometry-driven picture.

\begin{figure*}[t]
\centering
\begin{subfigure}[c]{0.34\textwidth}
    \includegraphics[width=\textwidth]{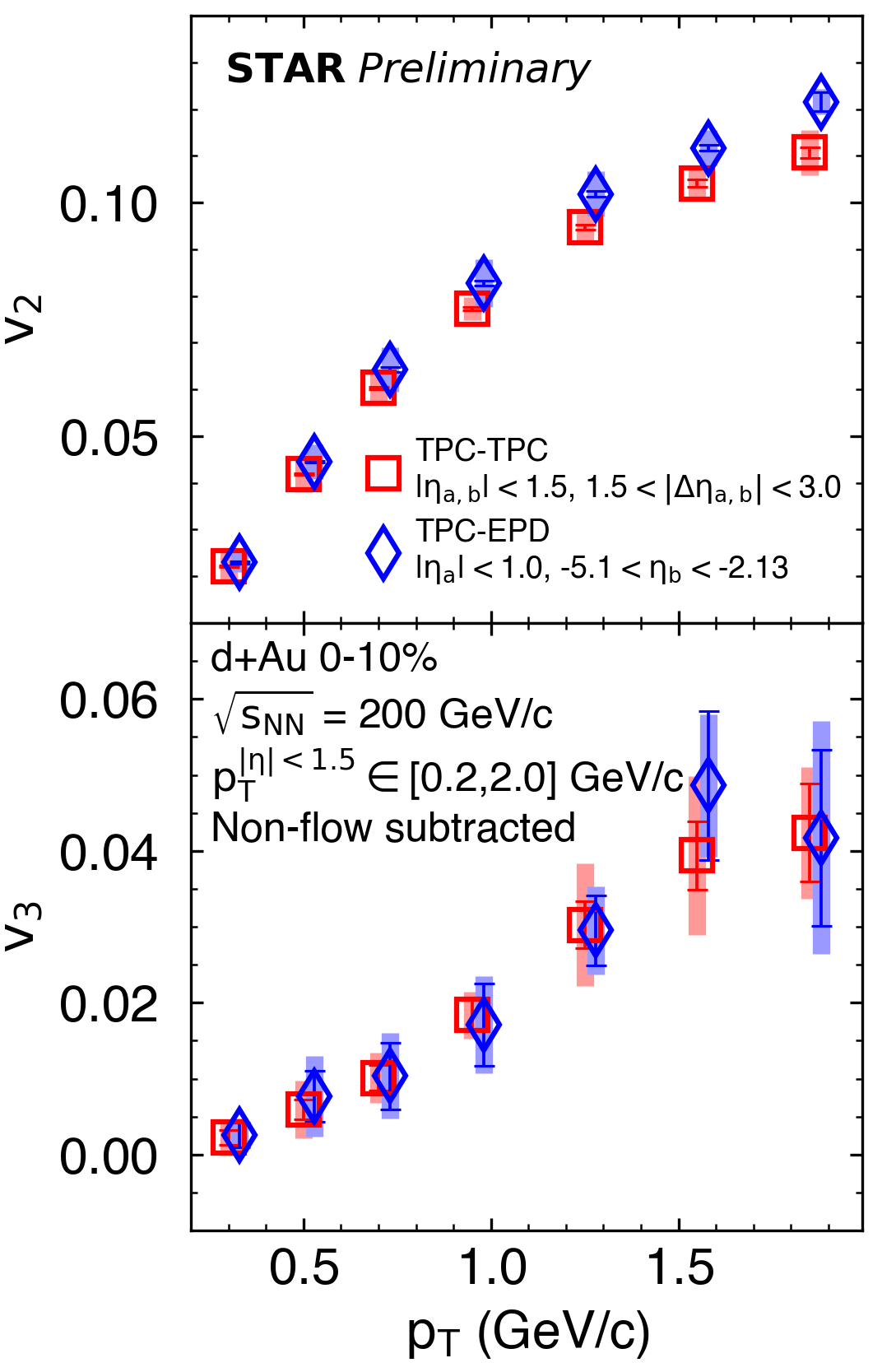}
    \caption{Comparison of $v_2$ (top) and $v_3$ (bottom) in central d+Au collisions measured with mid-mid (TPC-TPC) and mid-forward (TPC-EPD) correlations.}
    \label{fig:decorrelation}
\end{subfigure}
\hfill
\begin{subfigure}[c]{0.65\textwidth}
    \includegraphics[width=\textwidth]{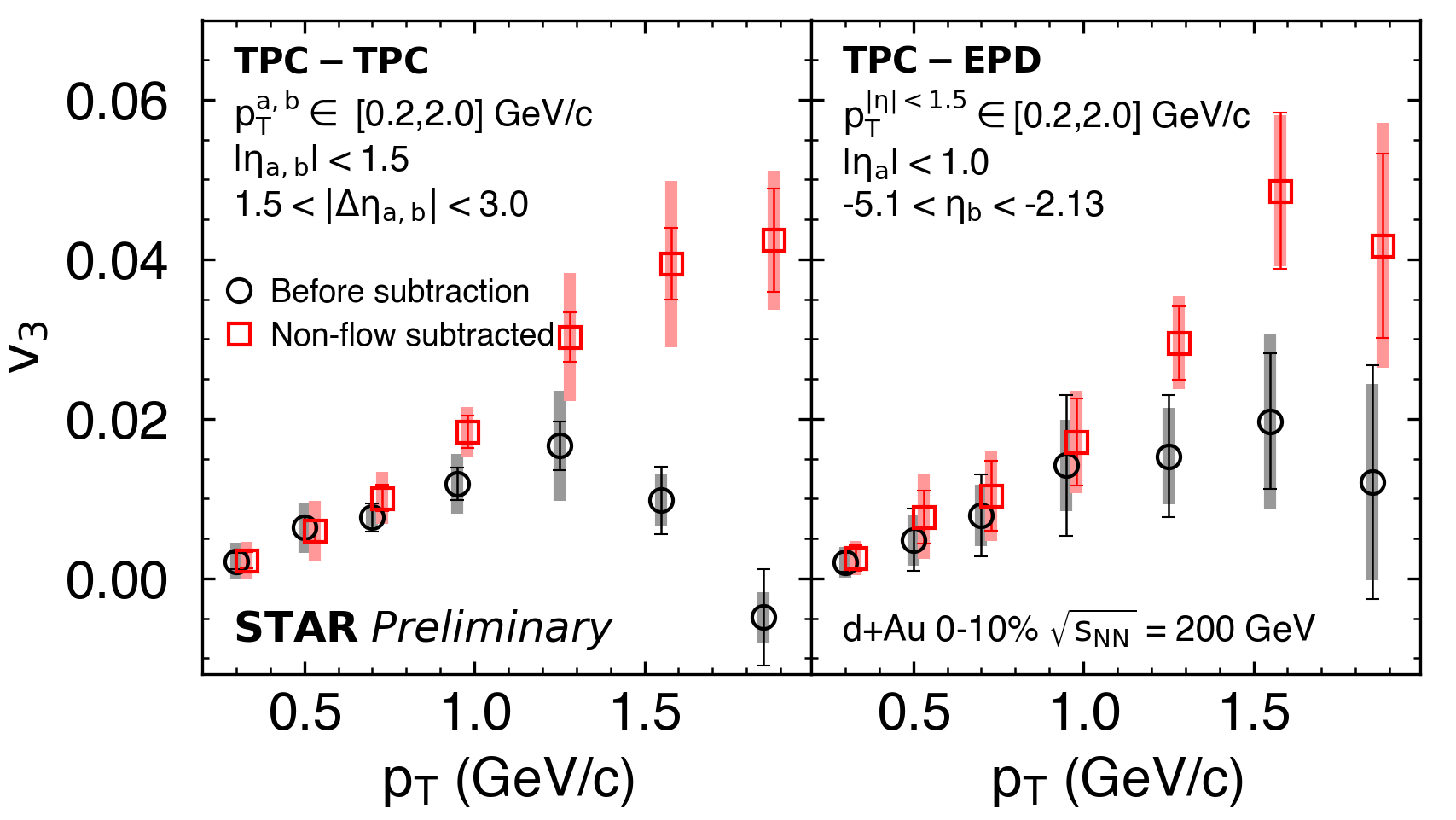}
    \caption{Illustration of non-flow effects on $v_3$ in central d+Au collisions using the subevent method. Results are shown before and after non-flow subtraction for both acceptances.}
    \label{fig:nonflow}
\end{subfigure}
\caption{Investigation of longitudinal decorrelation and non-flow effects in central d+Au collisions.}
\label{fig:decorrelation_nonflow}
\end{figure*}

Using the subevent method to compare mid-mid and mid-forward correlations, once non-flow contributions are subtracted, we find $v_n(\text{mid-mid}) \approx v_n(\text{mid-forward})$, as seen in Figure~\ref{fig:decorrelation}. This indicates that significant longitudinal decorrelation effects are not observed with this method. The analysis does, however, reveal that non-flow contributions are substantial even with a large rapidity gap and require careful subtraction, as shown in Figure~\ref{fig:nonflow}.

\section{Summary}
\label{sec-summary}
We have presented measurements of anisotropic flow in d+Au and $^{16}$O+$^{16}$O collisions at $\sqrt{s_{NN}} = 200$ GeV. The comparison provides strong evidence for a geometry-driven, collective origin of flow in small systems. The observed scaling of flow harmonics and their fluctuations with initial-state eccentricities supports the formation of a QGP droplet. The $v_3$ measurements specifically constrain the role of sub-nucleon fluctuations. Our investigation of correlations over a wide pseudorapidity range found no significant evidence for longitudinal flow decorrelation in the measured $v_n$s, but highlighted that non-flow effects are substantial and require careful handling.

\section*{Acknowledgements}
We thank the RHIC Operations Group and RCF at BNL. This work is supported by DOE Research Grant Number DE-SC0024602.

\bibliography{ref}

\end{document}